# P-N junctions dynamics in graphene channel induced by ferroelectric domains motion

Anatolii I. Kurchak[1], Eugene A. Eliseev[2], Sergei V. Kalinin[3], Maksym V. Strikha[1,4*], and Anna N. Morozovska[5†],

[1] *V.Lashkariov Institute of Semiconductor Physics, National Academy of Sciences of Ukraine, pr. Nauky 41, 03028 Kyiv, Ukraine*

[2] *Institute for Problems of Materials Science, National Academy of Sciences of Ukraine, Krjijanovskogo 3, 03142 Kyiv, Ukraine*

[3] *The Center for Nanophase Materials Sciences, Oak Ridge National Laboratory, Oak Ridge, TN 37831*

[4] *Taras Shevchenko Kyiv National University, Radiophysical Faculty pr. Akademika Hlushkova 4g, 03022 Kyiv, Ukraine*

[5] *Institute of Physics, National Academy of Sciences of Ukraine, pr. Nauky 46, 03028 Kyiv, Ukraine*

## Abstract

The p-n junctions dynamics in graphene channel induced by stripe domains nucleation, motion and reversal in a ferroelectric substrate is explored using self-consistent approach based on Landau-Ginzburg-Devonshire phenomenology combined with classical electrostatics. We revealed the extrinsic size effect in the dependence of the graphene channel conductivity on its length. For the case of perfect electric contact between the ferroelectric and graphene, relatively low gate voltages are required to induce the pronounced hysteresis of ferroelectric polarization and graphene charge in response to the periodic gate voltage. Pronounced nonlinear hysteresis of graphene conductance with a wide memory window corresponds to high amplitudes of gate voltage. We predict that the considered nano-structure "top gate/dielectric layer/graphene channel/ferroelectric substrate" can be a promising candidate for the fabrication of new generation of modulators and rectifiers based on the graphene p-n junctions.

---

* corresponding author, e-mail: maksym.strikha@gmail.com
† corresponding author, e-mail: anna.n.morozovska@gmail.com

# I. INTRODUCTION

Since the discovery of graphene [1, 2] its unique physical properties and application possibilities [3] have been attracting much attention of researchers. Following the development of graphene as a paradigm for 2D semiconductors, the remarkable properties of the p-n-junction (**pnJ**) in graphene have been realized experimentally [4, 5, 6] and studied theoretically [7, 8]. The first pnJ in graphene were realized by means of two gates doping one region with electrons and the other with holes [4-3]. Subsequently, pnJ was realized in a back gated graphene channel with a top gate superimposed on the dielectric layer above the channel [4]. For both cases, the electron transport through the pnJ potential barrier occurs in the direction normal to the barrier [7, 3], condition enabled by the fact that the electric field in the pnJ is typically high due to the low screening ability of two-dimensional Dirac quasiparticles and so the resistance of pnJ is quite low. The general physical principles of pnJs operation are formulated and explored by Beenakker [9]. Further experimental studies of pnJs in graphene were focused on the quantum Hall effect, the Klein (or Landau-Zener) tunneling and the Veselago lensing [10, 11].

Remarkably the usage of multiple gates doping of graphene channel by the opposite types of carriers is not the only possibility to design pnJs in graphene. The alternative promising and much less explored way was revealed till Hinnefeld et al [12] and Baeumer et al [13], who created a pnJ in graphene using the ferroelectric substrates $Pb(Zr,Ti)O_3$ and $LiNiO_3$, respectively. Further, Baeumer et al [13] considered the pnJ in graphene induced by a ferroelectric domain wall. The principal idea of the latter work is that if graphene is imposed on a $180^o$-ferroelectric domain wall, a pnJ can arise without applying any additional gates, doping or screening, due to the charge separation by an electric field of a ferroelectric domain wall − surface junction. The physical origin of pnJ appearance can be understood from Zheng et al [14] and Yusuf et al [15] results, who decided to combine ferroelectrics with graphene, since a pronounced free charge accumulation can take place at the graphene - ferroelectric interface [15]. Using the possibility of hysteretic ferroelectric gating, Zheng et al [14] demonstrated a symmetrical bit writing in graphene-on-ferroelectric FETs with an electro-resistance change of over 500%, as well as a reproducible, nonvolatile switching. Recently graphene-ferroelectric metadevices for nonvolatile memory and reconfigurable logic-gate operations have been proposed [16].

Kurchak and Strikha pointed out that adsorbed charges dynamics leads to the unusual conductivity effects in the graphene channel on an organic ferroelectric substrate [17]. Morozovska et al [18, 19] have shown that the effective screening of the spontaneous polarization discontinuity at the ferroelectric surface by graphene free charge strongly decreases the electrostatic energy of the graphene-on-ferroelectric and can cause versatile phenomena, such as the charge density modulation induced by the pyroelectric effect at the graphene/ferroelectric interface [18] and the triggering of the space charge modulation in multilayer graphene by ferroelectric domains [19]. Finite-size effects can strongly influence the nonlinear hysteretic dynamics of the stored charge and electro-resistance in the multilayer graphene-on-ferroelectric structures; in this case, the domain stripes of different polarities can induce domains with *p*- and *n*-type conductivity, and with pnJ potentials [20]. A theoretical model for the electric field and ballistic transport in a single-layer graphene channel at a 180°- ferroelectric domain wall have been developed recently [21].

The results of [21] have been expanded for the case of different types of current regimes in Ref.[22]. In [22] we presented the theory of the conductivity of pnJ in graphene channel, placed on ferroelectric substrate, caused by ferroelectric domain wall for the case of arbitrary current regime: from purely ballistic to diffusive one. We calculated the ratio of the pnJ conductions for opposite polarities of voltages, applied to source and drain electrodes of the channel, $G_+^{total}/G_-^{total}$, as the function of the graphene channel length *L*, electron mean free path λ and ferroelectric permittivity $\varepsilon_{33}^f$. We have demonstrated, that the small values of $G_+^{total}/G_-^{total}$ (0.1 and smaller), which correspond to efficient graphene pnJ based rectifier, can be obtained for the ferroelectrics with high permittivity ($\varepsilon_{33}^f \gg 100$) and for the ratios of *L*/λ~1 or smaller.

Here, complementary to Refs.[20-22], we consider the multi-domain states dynamics in a ferroelectric substrate. Using a self-consistent approach based on Landau-Ginzburg-Devonshire phenomenology combined with classical electrostatics, here we study p-n junctions dynamics in graphene channel induced by stripe domains nucleation, motion and reversal in a ferroelectric substrate. We regard that 2D-electrons in a single-layer graphene sheet have a linear Dirac density of states.

## II ANALYTICAL TREATMENT

**A. Problem statement**

The geometry of the considered problem is shown in the **Figs. 1(a).** Top gate is deposited on oxide layer; then 2D-graphene layer (channel) is separated from a ferroelectric substrate by ultra-thin paraelectric dead layer, originated due the imperfect deposition process of graphene on the ferroelectric. The ferroelectric substrate is in ideal electric contact with the bottom gate electrode. Periodic voltage is applied to the top gate. The voltage can induce 180-degree ferroelectric domain walls **(FDW)** motion in the ferroelectric substrate.

Schematic of the pnJ created in the graphene channel by domain walls moving in the ferroelectric substrate is shown in the **Figs. 1(b).** Since the lateral dimension of a ferroelectric film *L*<sub>FE</sub> is typically much higher than the graphene channel length *L*, odd, even or fractional number of domain walls can pass along the channel during the period of the gate voltage depending on the interrelation between the graphene channel length $L$ and the period $T_{FE}$ of the domain structure in a ferroelectric film.

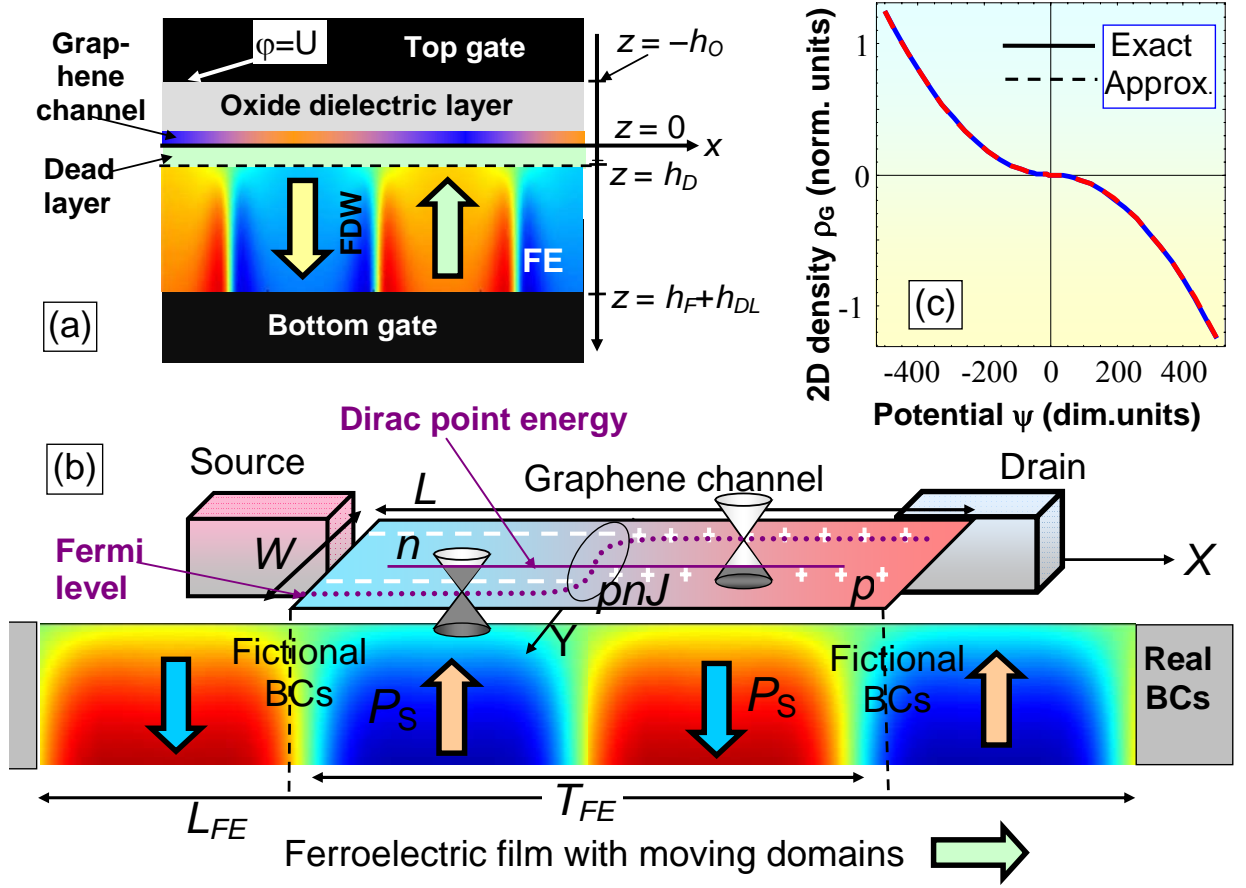

**FIG. 1.** **(a)** Schematics of the 180°-domain wall structure near the ferroelectric surface in the heterostructure "top gate – oxide dielectric layer – graphene channel – paraelectric dead layer – ferroelectric film – bottom gate". **(b)** Schematics of the pnJ induced in the graphene channel by domain walls moving in the ferroelectric substrate. **(c)** Dependence of the 2D charge density of graphene on the dimensionless variable $\psi = \dfrac{e\varphi + E_F}{k_B T}$. Solid curve is exact expression (1), dashed curve is Pade-exponential approximation (2).

Below we describe how we model each of the layers in the hetero-structure.

**Single-layer graphene channel.** We treat a single-layer graphene as an infinitely thin sheet for which that the two-dimensional (2D) electron density of states (DOS) is $g_n(\varepsilon) = g_p(\varepsilon) = 2\varepsilon/(\pi \hbar^2 v_F^2)$ (see e.g. [23]). Hence the 2D concentration of electrons in the conduction band and holes in the valence band of graphene are $n_{2D}(\varphi) = \int_0^\infty d\varepsilon\, g_n(\varepsilon) f(\varepsilon - E_F - e\varphi)$ and $p_{2D}(\varphi) = \int_0^\infty d\varepsilon\, g_p(\varepsilon) f(\varepsilon + E_F + e\varphi)$, respectively ($E_F$ is a Fermi energy level, these expressions correspond the gapless graphene spectrum). For pure planar graphene sheet we introduce a new variable, $\psi = \dfrac{e\varphi + E_F}{k_B T}$. Using the variable the graphene charge density $\sigma_G(\psi) = e(p_{2D}(\psi) - n_{2D}(\psi))$ is equal to

$$\sigma_G(\psi) = \dfrac{2(k_B T)^2 e}{\pi \hbar^2 v_F^2}\left(\text{Li}_2(-\exp(\psi)) - \text{Li}_2(-\exp(-\psi))\right) \qquad (1)$$

Here $\text{Li}_n(z) = \sum_{m=1}^{\infty} \dfrac{z^m}{k^n}$ is the polylogarithm function (see **Appendix A** of **Suppl. Mat.** [24]). Corresponding Pade-exponential approximation of Eq.(2) valid for arbitrary $\psi$ values is

$$\sigma_G(\psi) \approx \dfrac{2(k_B T)^2 e}{\pi \hbar^2 v_F^2}\left(\dfrac{1}{\eta(\psi)} - \dfrac{1}{\eta(-\psi)}\right), \qquad (2)$$

where the function: $\eta(\psi) = \exp(\psi) + 2\left(\psi^2 + \dfrac{\psi}{2} + \dfrac{2\pi^2}{12-\pi^2}\right)^{-1}$. Since the problem of dielectric permittivity of 2D-graphene layer is still under debate (see e.g. [25]), our results obtained for the DOS from Eqs.(1) are free from the problems inherent for 3D Boltzmann or Debye-Thomas-Fermi approximations [20, 21].

**Dielectric and dead layers.** Equations of state $\mathbf{D} = \varepsilon_0 \varepsilon_O \mathbf{E}$ and $\mathbf{D} = \varepsilon_0 \varepsilon_{DL} \mathbf{E}$ relate the electrical displacement $\mathbf{D}$ and electric field $\mathbf{E}$ in the oxide dielectric and ultrathin paraelectric dead layers of thicknesses $h_O$ and $h_{DL}$, respectively, $\varepsilon_0$ is a universal dielectric constant. The relative permittivity of the dead layer $\varepsilon_{DL}$ is rather high $\sim(10 - 10^2)$, in comparison with 1 for a physical gap. The potential $\varphi_{DL}$ satisfies Laplace's equation inside the dead layer.

**Ferroelectric substrate.** As a substrate we consider a ferroelectric film of thickness $l$ with ferroelectric polarization $P_3^f$ directed along its polar axis z, with 180-degree domain wall – surface junctions [see **Fig. 1(a)**]. Also we assume that the dependence of polarization components on the inner field $\mathbf{E}$ can be linearized for transverse components as $P_1 = \varepsilon_0(\varepsilon_{11}^f - 1)E_1$ and $P_2 = \varepsilon_0(\varepsilon_{22}^f - 1)E_2$. Ferroelectric is dielectrically isotropic in transverse directions, i.e. relative dielectric permittivity are equal, $\varepsilon_{11}^f = \varepsilon_{22}^f$. Polarization z-component is $P_3(\mathbf{r}, E_3) = P_3^f(\mathbf{r}, E_3) + \varepsilon_0(\varepsilon_{33}^b - 1)E_3$, where a relative background permittivity $\varepsilon_{ij}^b \leq 10$ is introduced [26]. The ferroelectric permittivity $\varepsilon_{33}^f \gg \varepsilon_{33}^b$. Inhomogeneous spatial distribution of the ferroelectric polarization $P_3(x, y, z)$ will be determined from the time-dependent Landau-Ginzburg-Devonshire (LGD) type Euler-Lagrange equation,

$$\Gamma \dfrac{\partial P_3}{\partial t} + aP_3 + bP_3^3 + cP_3^5 - g\Delta P_3 = E_3. \qquad (3a)$$

$\Gamma$ is a Landau-Khalatnikov relaxation coefficient, and $g$ is a gradient coefficient, $\Delta$ stands for a 3D-Laplace operator. Corresponding boundary conditions are of the third kind [27],

$$\left(P_3 - \Lambda_+ \dfrac{\partial P_3}{\partial z}\right)\bigg|_{z=W} = 0, \qquad \left(P_3 + \Lambda_- \dfrac{\partial P_3}{\partial z}\right)\bigg|_{z=h} = 0 \qquad (3b)$$

The physical range of extrapolation lengths $\Lambda_\pm$ is (0.5 – 2) nm [28]. Constants $a = \alpha_T(T - T_C)$, $b$ and $c$ are the coefficients of LGD potential expansion on the polarization powers (also called as linear and nonlinear dielectric stiffness coefficients). Quasi-static electric field is defined via electric potential as $E_3 = -\partial\varphi/\partial z$. The potential $\varphi_f$ satisfies Poisson equation inside a ferroelectric film.

Also we suppose that the polarization relaxation time $\tau_{LK} = \Gamma/|a_3| \sim 10^{-11}$ s is much higher than the graphene charge relaxation time $\tau_M = \varepsilon\varepsilon_0/(e\eta n_s) \sim 10^{-12}$ s ($\varepsilon_0$ is a universal dielectric constant, $\varepsilon$ is a relative dielectric permittivity, $\eta$ is a mobility of carriers in graphene), and so adiabatic approximation can be used for the charge description [see **Fig.S1** in [24]]. The inequality $\tau_{LK} \gg \tau_M$ works well in the vicinity of ferroelectric phase transition, where $\tau_{LK}$ diverges due to the critical slowing down effect.

Hence, for the problem geometry shown in the **Fig. 1(a)** the system of electrostatic equations acquires the form:

$$\Delta\varphi_O = 0, \quad \text{for} \quad -h_O < z < 0, \quad \text{(oxide dielectric layer "}O\text{")} \quad (4a)$$

$$\Delta\varphi_{DL} = 0, \quad \text{for} \quad 0 < z < h_{DL}, \quad \text{(dead layer "}DL\text{")} \quad (4b)$$

$$\left(\varepsilon_{33}^b \frac{\partial^2}{\partial z^2} + \varepsilon_{11}^f \Delta_\perp\right)\varphi_f = \frac{1}{\varepsilon_0}\frac{\partial P_3^f}{\partial z}, \quad \text{for} \quad h_{DL} < z < h_{DL} + h_F. \quad \text{(ferroelectric "}f\text{")} \quad (4c)$$

3D-Laplace operator is $\Delta$, 2D-Laplace operator is $\Delta_\perp$. Boundary conditions to the system (4) are fixed potential at the top ($z = -h_O$) and bottom ($z = h_{DL} + h_F \approx h_F$) gate electrodes; the continuity of the electric potential at the graphene layer ($z = 0$) and the equivalence of difference of the electric displacement normal components, $D_3^O = \varepsilon_0\varepsilon_O E_3$ and $D_3^{DL} = \varepsilon_0\varepsilon_{DL}E_3$, to the surface charges in graphene $\sigma_G(x, y)$; and the continuity of the displacement normal components, $D_3^f = \varepsilon_0\varepsilon_{33}^b E_3 + P_3^f$ and $D_3^{DL} = \varepsilon_0\varepsilon_{DL}E_3$, at dead layer/ferroelectric interface.

Explicit form of the boundary conditions (**BCs**) in z–direction is:

$$\varphi_O(x, y, -h_O) = U(t), \quad (5a)$$

$$\varphi_O(x, y, 0) = \varphi_{DL}(x, y, 0), \quad D_3^O(x, y, 0) - D_3^{DL}(x, y, 0) = \sigma_G(x, y), \quad (5b)$$

$$\varphi_d(x, y, h_{DL}) = \varphi_f(x, y, h_{DL}), \quad D_3^{DL}(x, y, h_{DL}) - D_3^f(x, y, h_{DL}) = 0, \quad (5c)$$

$$\varphi_f(x, y, h_{DL} + h_F) = 0. \quad (5d)$$

The gate voltage is periodic with a period $T_g$, $U(t) = U_{max}\sin(2\pi t/T_g)$.

To generate moving domains ferroelectric film thickness should be above the critical thickness $l_{cr}$ of the size-induced phase transition into a paraelectric phase. The critical thickness of the single-domain ferroelectric state instability strongly depends on the dielectric and dead layer thicknesses, namely [29], $l_{cr}(T) \approx \dfrac{1}{\alpha_T(T_C - T)} \left( \dfrac{2g}{\lambda + L_C} + \dfrac{h_{DL}}{\varepsilon_0 \varepsilon_{DL}} + \dfrac{h_O}{\varepsilon_0 \varepsilon_O} \right)$, where $L_C = \sqrt{g/\varepsilon_0 \varepsilon_{33}^b} \sim 0.1$ nm is a correlation length in z-direction that is very small due to the depolarization field effect [30, 31].

**B. The "fictional" boundary conditions at the source and drain planes**

The domains origin can be energetically favorable above the critical thickness, since they minimize the depolarization field energy in the gap and dielectric layer [32]. The period of domain stripes depends on the thicknesses of oxide dielectric layer $h_O$, dead layer $h_{DL}$ and film $h_F$ in a self-consistent way.

We note that exact domain dynamics will be determined by the interplay of the driving forces, memory effects in materials due to defects, and parity effects. Here, we consider only the latter, assuming no pinning on the top surface. Since the lateral dimension of a ferroelectric film $L_{FE}$ is typically much higher than the graphene channel length $L$ in real experiments, ***odd, even*** or ***fractional*** number of domain walls can pass along the channel, expand or contract in ferroelectric during the period of the gate voltage $T_g$ depending on the interrelation between the graphene channel length $L$ and the period $T_{FE}$ of the domain structure in a ferroelectric film [**Fig.1(b)**].

Taking into account that $L_{FE} \gg L$, it is possible to perform calculations on a computational cell with a transverse dimension $-L/2 \leq x \leq L/2$. Thus we modelled the realistic situations of domain walls motion by imposing "fictional" ***periodic, antiperiodic*** or ***mixed parity*** boundary conditions (**BCs**) on polarization component $P_3\left(\pm \dfrac{L}{2}, y, z\right)$, its derivative $\left.\dfrac{\partial P_3}{\partial x}\right|_{x=\pm \frac{L}{2}}$, electric potential $\varphi_f\left(\pm \dfrac{L}{2}, y, z\right)$ and its derivative $\left.\dfrac{\partial \varphi_f}{\partial x}\right|_{x=\pm \frac{L}{2}}$ at the lateral boundaries $x = \pm L/2$ of the computation box. Mathematical forms of the *periodic*, *antiperiodic* or *mixed parity* BCs are listed in **Appendix B** of **Suppl. Mat.** [24], the physical consequences of the BCs are described below.

The *periodic*, *antiperiodic* or *mixed parity* BCs imposed on polarization and electric potential in transverse x-direction model the appearance and motion of different parity (*even* or *odd*) of pnJs along the channel of length $L$ over a voltage period, respectively. It appeared that periodic BCs for polarization and potential, allow the motion of the even number of domain walls in the ferroelectric and so the even number of pnJs are moving in the channel. Thus the asymmetry of the graphene channel conductance and rectification effect is absent in this case for the reasons discussed below. In contrast the mixed parity BCs, can lead to the asymmetry of the graphene channel conductance and rectification effect, which will be demonstrated in the next section. The completely antiperiodic BCs can lead to the motion of the odd number of domain walls in the ferroelectric and hence the odd number pnJs can move in the channel. The asymmetries of the graphene channel conductance and rectification effect are possible in the case.

One can try to understand the effect of the BCs on the graphene conduction kinetics induced by the moving domain walls using the analytical results [21, 22] for a thermodynamic state an instant "snapshot" at some moment of time. Because of the results [21, 22], it is natural to expect that for the *even* number (2k) of walls between source and drain electrodes of graphene channel its conductions $G_+^{total}$ and $G_-^{total}$ are the same of the for both polarities of the gate voltage

$$\frac{G_+^{total}}{G_-^{total}} = 1, \quad (6a)$$

because for each polarity there are *k* pnJ with conduction, given by eq.(8) of [22] (electrons are tunneling trough the barrier), and *k* pnJ with conduction, given by eq.(10) of [22] (electrons do not feel the barrier). For the *odd* number of the walls 2k+1, Eq.(14) of Ref.[22] can be transformed as:

$$\frac{G_+^{total}}{G_-^{total}} = \frac{\beta(L+\lambda) + \lambda k(1+\beta)}{\beta(L+\lambda) + \lambda k(1+\beta) + \lambda}, \quad (6b)$$

where the factor β is equal to $\beta = \sqrt{\frac{\pi\alpha c}{4\varepsilon_{33}^f v_F}}$, *L* is the graphene channel length, λ is the electron mean free path, $\varepsilon_{33}^f$ is the ferroelectric permittivity, $v_F$ is the Fermi velocity of electrons in graphene, α = 1/137 is fine structure constant, *c* is light velocity in vacuum.

For a pronounced diffusion regime of current, $\beta L \gg \lambda$, and/or for the great number of the walls, $k \gg k_{cr}$, where $k_{cr} = 1 + \frac{\beta L}{(1+\beta)\lambda}$, the right-hand-side in Eq.(6b) tends to unity (as in Eq. (9a)), and a conduction of graphene channel is described by a well known expression [8]:

$$G = \frac{\lambda(n_{2D})}{L} \frac{2e^2}{\hbar\pi^{3/2}} w\sqrt{n_{2D}} \quad (7)$$

Here *w* is graphene channel width, $n_{2D}$ is 2D charge concentration in the channel, determined by Eq.(1). The mean free path $\lambda(n_{2D})$ corresponds to the concentration $n_{2D}$. For the scattering at ionized centres in the substrate in the temperature range far from Curie temperature, $\lambda(n_{2D}) \sim \sqrt{n_{2D}}$ [8]. Therefore the conduction is a linear function of concentration in this case.

## III. NUMERICAL RESULTS AND DISCUSSION

Below we present results of numerical modeling of the problem (1)-(5). We study numerically the modulation of the graphene channel conductance caused by a domain structure moving in a ferroelectric substrate. Parameters used in the calculations are listed in **Table I.**

**Table I**. Parameters designations and numerical values

| Parameter, constant or value | Numerical value and dimensionality |
|---|---|
| oxide dielectric thickness | $h_O = (4-10)$ nm |
| dielectric (dead) layer thickness | $h_{DL} = (0-4) \times 10^{-10}$ m |
| ferroelectric film thickness | $h_F = (50-500)$ nm |
| graphene channel length | $L = (20-2000)$ nm |
| universal dielectric constant | $\varepsilon_0 = 8.85 \times 10^{-12}$ F/m (e/Vm) |
| permittivity of the dielectric layer | $\varepsilon_{DL} = 100$ (typical range $10-300$) |
| permittivity of the ferroelectric film | $\varepsilon_{33}^f = 500$, $\varepsilon_{11}^f = \varepsilon_{22}^f = 780$ (Pb(ZrTi)O$_3$-like) |
| Landau-Ginzburg-Devonshire potential coefficients | $\alpha_T = 2.66 \times 10^5$ C$^{-2}$·mJ/K, $T_C = 666$ K (PbZr$_x$Ti$_{1-x}$O$_3$ (x ≈ 0.5)), $P_S^{bulk} = (0.5 - 0.7)$ C/m$^2$ $b = 1.91 \times 10^8$ J C$^{-4}$·m$^5$, $c = 8.02 \times 10^8$ J C$^{-6}$·m$^9$ |
| dielectric anisotropy of ferroelectric film | $\gamma = \sqrt{\varepsilon_{33}^f/\varepsilon_{11}^f} = 0.8$ |
| extrapolation length | $\Lambda_+ = \Lambda_- = \infty$ |
| Plank constant | $\hbar = 1.056 \times 10^{-34}$ J·s $= 6.583 \times 10^{-16}$ eV·s |
| Fermi velocity of electrons in graphene | $v_F \approx 10^6$ m/s |

For the enlisted parameters we get factor $\beta = 0.185$, the product $\beta L \approx (4 - 400)$ nm for the channel length $L=(20 - 2000)$ nm, and $k_{cr} = 1 + 0.16L/\lambda$. So that the ratio $G_+^{total}/G_-^{total}$ can be noticeably different from unity for $\lambda \geq 0.185L$ and domain wall number $k \leq 1 + 0.16L/\lambda$.

The polarization component in the ferroelectric film, $P_3$, variation of 2D-concentration of free carriers in the graphene channel, $\Delta n_G = (p_{2D} - n_{2D})$, and the effective conductance ratio $\Delta\eta(U_{max}) = \frac{\Delta n_G(+U_{max})}{\Delta n_G(-U_{max})}$ were calculated in dependence on the gate voltage $U(t) = U_{max} \sin(2\pi t/T_g)$. The maximal charge density $\sigma_G = en_G \sim 0.05$ C/m$^2$ recalculated from the $\Delta n_G \sim 10^{17}$ m$^{-2}$ appeared noticeably smaller than the spontaneous polarization of the ferroelectric film $P_S \sim 0.5$ C/m$^2$ because of the electric potential drop in the oxide dielectric layer and corresponding appearance of depolarization field in it.

The hysteresis loops of the average polarization $P_3(U)$ and total concentration variation $\Delta n_G(U)$ are shown in **Figs.2(a)-(f).** Only the steady regime is shown, for all the cases the transient process was cut off. Black, red and magenta loops correspond to the different amplitudes of gate voltage $U_{max} = (2, 5, 10)$ V. At relatively low voltages (≤ 2 V for chosen material parameters) polarization and concentration loops of quasi-elliptic shape do not reveal any ferroelectric peculiarities [see black curves in **Fig. 2(a)-(f)**]. This happens because the film state is poly-domain, and the domain walls do not disappear with the gate voltage changing from $-U_{max}$ to $+U_{max}$, but they are moved by the electric field. Actually, the domain walls are moving to minimize the system energy when the voltage higher than the coercive one is applied to the top gate. Note that the domain wall pinning effects which can affect on the coercive field are not considered in our modeling.

With the increase of the gate voltage amplitude $U_{max}$ to (5 – 10) V the domain walls start to collide and the domains with opposite polarization orientation almost "annihilate" for definite periodic moments of time $t$, and then the film polar state with some degree of unipolarity partially restores [see red and magenta curves in **Figs.2(a)-(c)** and **Suppl. Movies A-C**].

The degree of unipolarity of the ferroelectric film strongly increases with $U_{max}$ increase to 15 V and for $U_{max} \geq$ (15 – 25) V we observed that the stationary switching regime is single-domain independently on the poly-domain seeding [see blue and magenta curves in **Figs. S3** in Ref.[24]]. Note that for the voltages higher than 10 V (depending on the BCs type) one needs especially good quality of dielectric layers, because the field in the oxide layer can be greater than the breakdown one (about 1 V/nm for $SiO_2$ [33]).

The loops asymmetry and rectification ratio is defined by the symmetry of the BCs. In particular, completely symmetric loops of polarization $P_3(U)$ and concentration $\Delta n_G(U)$ variation correspond to the periodic BCs [**Figs.2(a)** and **2(c)**]. Positive (holes) and negative (electrons) charges amount are equal entire the period of the gate voltage for the periodic BCs. The rectification effect is absent in the case of periodic BCs, since the effective ratio $\Delta \eta(U_{max}) \equiv -1$ for all $U_{max}$ [see **Figs.2(g)**], because the even number of domain walls is moving in the ferroelectric substrate at any time [see **Suppl. Movie A** in Ref.[24]]. A typical distributions of electric potential and polarization proved that two domain walls separating three domains with "up" and "down" polarizations are present in the ferroelectric film [see **Fig.3(a)** and **Fig.S2(a)** in Ref.[24]]. An example of the spatial-temporal evolution of the free charge concentration calculated along the graphene channel is shown in **Figs. 3(d)**. Two pnJs produced by the two moving domain walls are seen at certain times over one period. The relative depth (i.e. maximum to minimum ratio) and position ($\Delta n_G = 0$) of the pnJs are time-dependent.

Asymmetric loops of $P_3(U)$ and $\Delta n_G(U)$ correspond to the mixed BCs [**Figs.2(b)** and **2(e)**]. The vertical asymmetry and horizontal shift of polarization loop are much stronger than the asymmetry of concentration variation because the polarization acts on the charge indirectly via the depolarization field. The asymmetry of $P_3(U)$ and $\Delta n_G(U)$ becomes weaker with maximal voltage increase [compare different loops in **Figs.2(b)** and **2(e)**]. The rectification effect is present in the case of low voltages, since the effective ratio $\Delta \eta(U_{max}) \gg 1$ for $1 < U_{max} < 3$, the ratio changes sign at $U_{max} = 4$, and then $\Delta \eta(U_{max}) \to -1$ for higher voltages [see **Figs.2(h)**], because asymmetric domain walls are moving in the ferroelectric film [see **Suppl. Movie B** in Ref.[24]]. An instant distributions of electric potential and polarization show that two domain walls with complex shape near the surface, which separate three domains (two with "up" and one with "down" polarizations), are located in the ferroelectric film [see **Fig.3(b)** and **Fig.S2(b)** in Ref.[24]]. Corresponding spatial-temporal evolution of the free charge concentration calculated along the graphene channel is shown in **Figs. 3(e).** Two pnJs produced by the two moving domain walls exist inside the channel at all times over one period. Also there are two halves of "positive" pnJs at the channel boundaries and, most probably, these peculiarities create the asymmetry of the total conductance.

The antiperiodic BCs also lead to asymmetric loops of polarization $P_3(U)$ and concentration $\Delta n_G(U)$ [**Figs.2(c)** and **2(f)**]. As well as in the previous case, the vertical asymmetry and horizontal shift of $P_3(U)$ loop are much stronger than the asymmetry of $\Delta n_G(U)$ loop because the polarization acts on the charge indirectly via the depolarization field. The asymmetry of $P_3(U)$ and $\Delta n_G(U)$ becomes weaker with maximal voltage increase [compare different loops in **Figs.2(c)** and **2(f)**]. The rectification effect is evident in the case of low and moderate voltages, since the effective ratio $\Delta\eta(U_{max}) \ll 1$ for $1 < U_{max} < 4$, and for higher voltages then the ratio saturates, $\Delta\eta(U_{max}) \to -1$ [see **Figs.2(i)**], because the odd number of domain walls is moving in the ferroelectric film most of the time [see **Suppl. Movie C** in Ref.[24]]. A typical distributions of electric potential and polarization show that three domain walls, which separate four domains (two with "up" and two with "down" polarizations), are present in the ferroelectric film [see **Fig.3(c)** and **Fig.S2(c)** in Ref.[24]]. A typical spatial-temporal evolution of the graphene charge is shown in **Figs. 3(f).** Three pnJs produced by the three moving domain walls exist at almost all times over one period [see **Suppl. Movie C** in Ref.[24]], at that their relative depth and position are time-dependent.

Spatial-temporal distributions of 2D-concentration of the free charge $\Delta n_G(x,t)$ calculated for the small amplitude of the gate voltage ($U_{max} = 2$ V) for the periodic, mixed and antiperiodic BCs are shown in **Fig.4(a), (b)** and **(c),** respectively. Maximal positive (holes) and negative (electrons) charge densities are equal entire the period of the gate voltage for the periodic BCs [corresponding color scale of the carriers density ranges from $+8.7\times10^{17}$ m$^{-2}$ to $-8.7\times10^{17}$ m$^{-2}$ in **Fig.4(a)**], reflecting the fact that negative and positive polarizations are equivalent entire the periodic for periodic BCs [see symmetric polarization loops in **Figs.2(a)**]. The charge density symmetry is broken for the mixed BCs [corresponding color scale ranges from $+3.6\times10^{17}$ m$^{-2}$ to $-4.4\times10^{17}$ m$^{-2}$ in **Fig.4(b)**], reflecting the fact that polarization has noticeable degree of unipolarity for the mixed BCs [see asymmetric polarization loops in **Figs.2(b)**]. The charge density symmetry breaking for the antiperiodic BCs is evident from **Fig.4(c)**, where the carrier density scale ranges from $+3.7\times10^{17}$ m$^{-2}$ to $-4.2\times10^{17}$ m$^{-2}$, caused by the polarization asymmetry shown in **Figs.2(c)**. To resume, the electric boundary conditions for polarization at the lateral surfaces of ferroelectric substrate rules the asymmetry of the graphene channel conductance between the source and drain electrodes.

We note the pronounced extrinsic size effect in the dependence of the graphene channel conductivity on its length. Corresponding contour maps in coordinates "channel length – time" were calculated for periodic BCs and shown in **Figs.5**

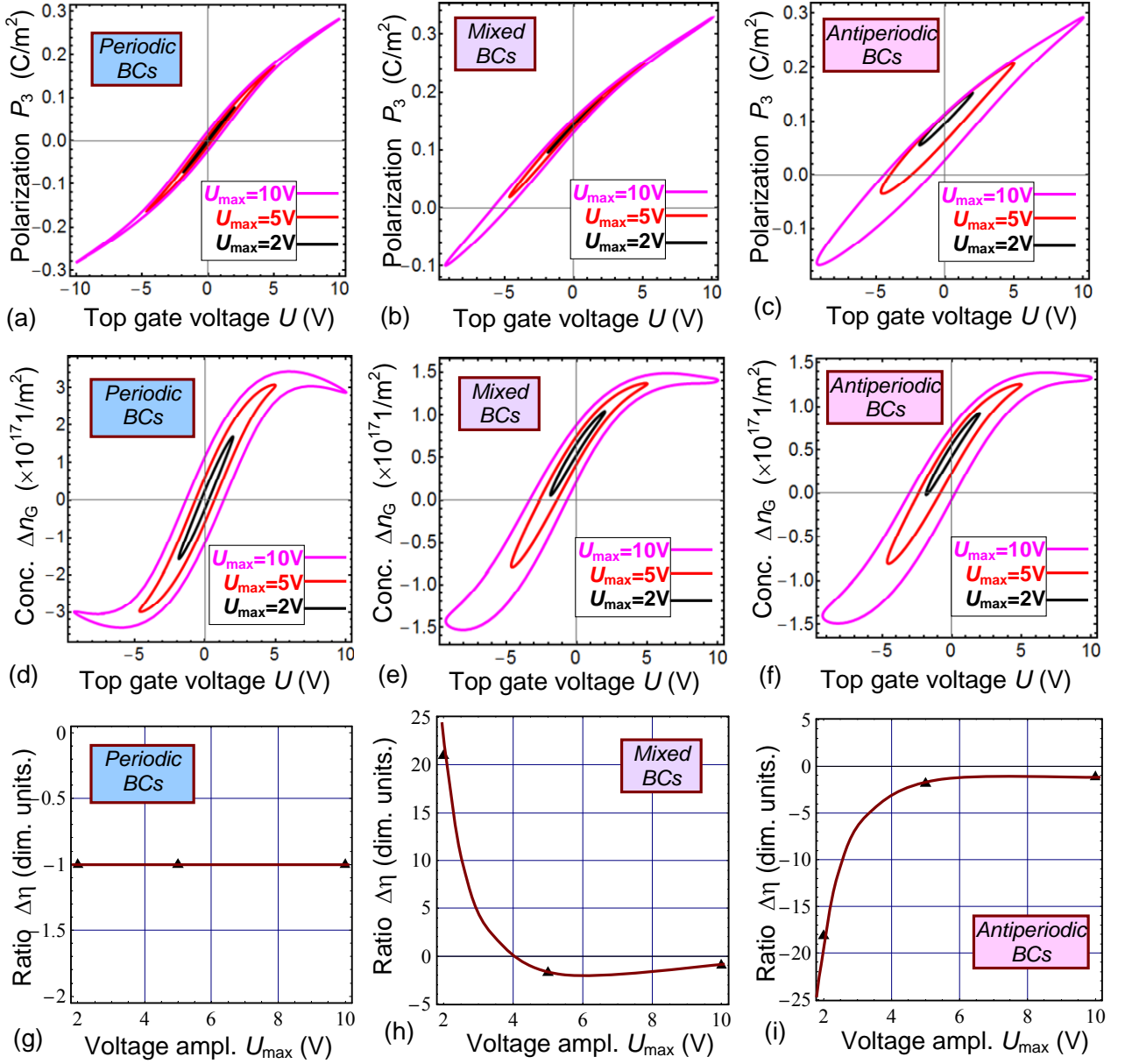

**FIG. 2.** Hysteresis loops of ferroelectric polarization $P_3(U)$ **(a, b, c)**, carriers concentration variation in graphene channel $\Delta n_G(U)$ **(d, e, f)** and the conductance ratio $\Delta\eta(U_{max})$ **(g, h, i)** calculated for periodic **(a, d, g)**, mixed **(b, e, h)** and antiperiodic **(c, f, i)** boundary conditions (BCs). Black, red and magenta loops in plots **(a)-(f)** correspond to the different amplitudes of gate voltage $U_{max}$ = (2, 5, 10) V. The thicknesses of ferroelectric film is $h_f$=75 nm, oxide dielectric thickness $h_O$=8 nm, dead layer thickness $h_{DL}$=0.4 nm, gate voltage period $T_g$=10$^3$ s. Other parameters are listed in **Table I.**

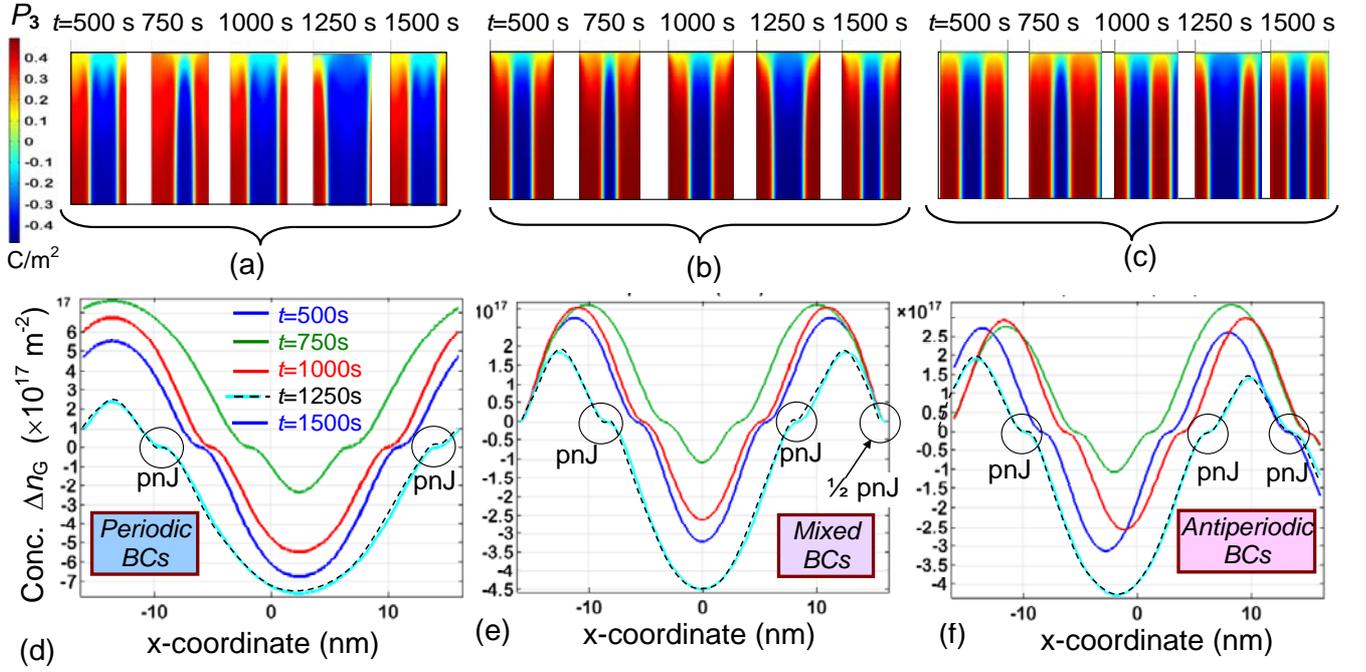

**FIG. 3.** Spatial distribution of polarization component $P_3$ in a ferroelectric film calculated at certain times over a period, 500, 750, 1000, 1250 and 1500 s (specified in the plots) for the periodic **(a)**, mixed **(b)** and antiperiodic **(c)** BCs. 2D-concentration of the free charge $\Delta n_G(x,t)$ calculated along the graphene channel at certain times over a period, 500, 750, 1000, 1250 and 1500 s (specified in the plots) for the periodic **(d)**, mixed **(e)** and antiperiodic **(f)** BCs. Gate voltage amplitude $U_{max} = 5$ V and period $T_g = 1000$ s. Other parameters are the same as in **Fig.2.** The transient process that almost vanish enough rapidly is not shown (so that we started to show the plots from t=500 s).

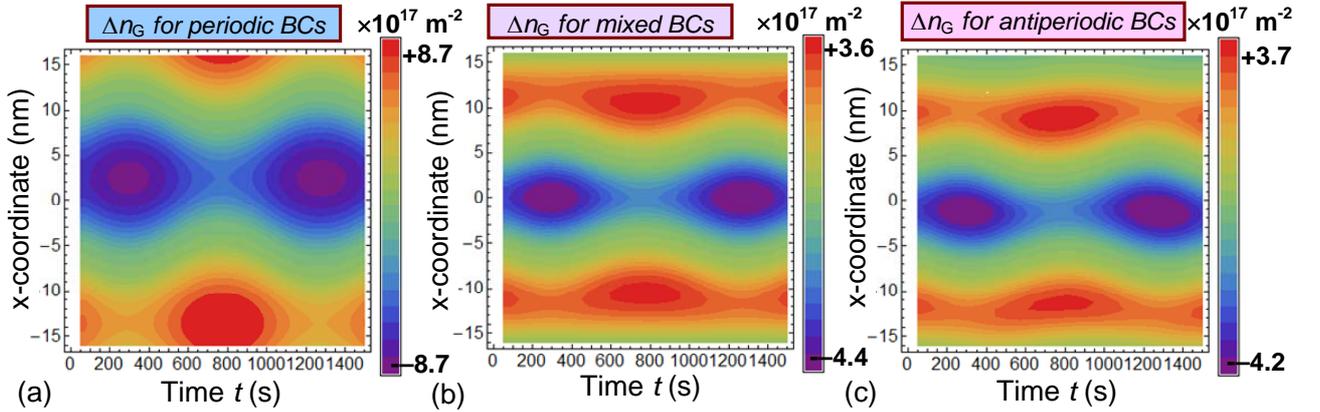

**FIG. 4.** Spatial-temporal distributions of 2D-concentration of the free charge $\Delta n_G(x,t)$ calculated along the graphene channel for the periodic **(a)**, mixed **(b)** and antiperiodic **(c)** BCs. Gate voltage amplitude $U_{max} = 2$ V and period $T_g = 1000$ s. Other parameters are the same as in **Fig.2.**

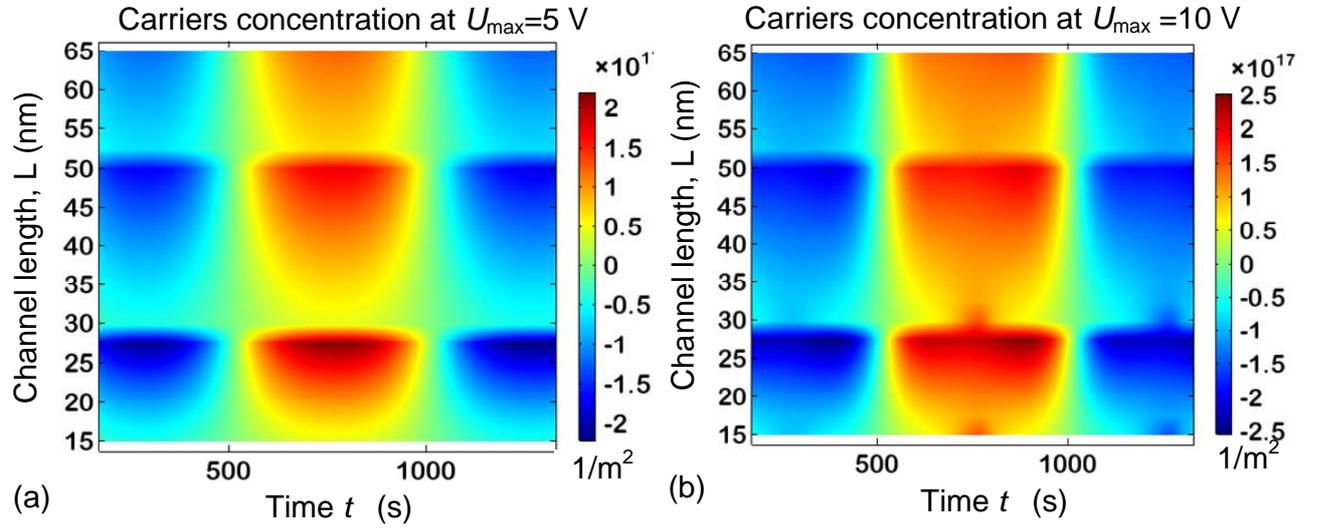

**FIG. 5.** Dependence of the graphene channel conductivity on its length and time calculated for the periodic BCs at $U_{max}$ = 5 V **(a)** and $U_{max}$ = 10 V **(b)**. Other parameters are the same as in **Fig.4**.

To resume results of this section we predict that the considered nano-structure "top gate/dielectric layer/graphene channel/ferroelectric substrate" can be a promising candidate for the fabrication of new generation of modulators and rectifiers based on the graphene p-n junctions.

## IV. CONCLUSION

Using a self-consistent approach based on Landau-Ginzburg-Devonshire phenomenology combined with classical electrostatics we studied p-n junctions dynamics in graphene channel induced by stripe domains nucleation, motion and reversal in a ferroelectric substrate. For the case of intimate electric contact between the ferroelectric and graphene sheet relatively low gate voltages are required to induce the pronounced hysteresis of ferroelectric polarization and graphene charge in dependence on the periodic gate voltage. The harmonic-like charge concentration modulation caused by a domain structure alternates with its homogeneous distribution caused by a single-domain states under the periodic change of applied voltage.

Pronounced nonlinear hysteresis of graphene conductance corresponds to high amplitudes of gate voltage. Nevertheless electric potential drops in the dielectric layer and gap, the graphene charge recalculated from the carriers concentration appeared of the same order than the spontaneous polarization of the ferroelectric film. This happens because the gate voltages are high enough to increase the polarization of the ferroelectric film entire the period.

We established the electric boundary conditions for polarization at the lateral surfaces of ferroelectric substrate rules the asymmetry of the graphene channel conductance between the source and drain electrodes. Also we revealed the pronounced extrinsic size effect in the dependence of the graphene channel conductivity on its length.

We predict that the considered nano-structure "top gate/dielectric layer/graphene channel/ferroelectric substrate" can be promising candidates for the fabrication of new generation of modulators and rectifiers based on the graphene p-n junctions.

**Acknowledgements.** Authors are very grateful to Dr. Qian Li for useful suggestions. A.I.K. and A.N.M. acknowledges the State Fund of Fundamental Research of Ukraine. S.V.K. acknowledges the Office of Basic Energy Sciences, U.S. Department of Energy. Part of work was performed at the Center for Nanophase Materials Sciences, which is a DOE Office of Science User Facility.

**Authors' contribution.** A.I.K. and E.A.E. performed numerical calculations and make figures. M.V.S. generated the research idea and motivation, derived analytical expression for graphene conductance, strongly contributed to the results analyses and manuscript improvement. S.V.K. contributed to the interpretation of the results and manuscript writing. A.N.M. formulated the problem mathematically, performed most of analytical calculations, select parameters and wrote the manuscript.

SUPPLEMENTARY MATERIALS TO THE MANUSCRIPT

# P-N junctions dynamics in graphene channel induced by ferroelectric domains motion

Anatolii I. Kurchak[1], Eugene A. Eliseev[2], Sergei V. Kalinin[3], Maksym V. Strikha[1,4], and Anna N. Morozovska[5],

[1] *V.Lashkariov Institute of Semiconductor Physics, National Academy of Sciences of Ukraine, pr. Nauky 41, 03028 Kyiv, Ukraine*

[2] *Institute for Problems of Materials Science, National Academy of Sciences of Ukraine, Krjijanovskogo 3, 03142 Kyiv, Ukraine*

[3] *The Center for Nanophase Materials Sciences, Oak Ridge National Laboratory, Oak Ridge, TN 37831*

[4] *Taras Shevchenko Kyiv National University, Radiophysical Faculty pr. Akademika Hlushkova 4g, 03022 Kyiv, Ukraine*

[5] *Institute of Physics, National Academy of Sciences of Ukraine, pr. Nauky 46, 03028 Kyiv, Ukraine*

# APPENDIX A. Derivation of the charge density

We treat a single-layer graphene as a sheet of thickness $d$ of 3-5 Å order, for which that the electronic density of states (DOS) is $g(\varepsilon) = 2\varepsilon/(\pi\hbar^2 v_F^2)$ (see Ref in the main text). Hence the two-dimensional (2D) concentration of electrons in the conduction band is:

$$n_S(\varphi) = \int_0^\infty d\varepsilon\, g_n(\varepsilon) f(\varepsilon + E_C - E_F - e\varphi) \equiv \int_0^\infty \frac{2\varepsilon d\varepsilon/(\pi\hbar^2 v_F^2)}{1 + \exp((\varepsilon + E_C - E_F - e\varphi)/k_B T)}$$
$$= -\frac{2(k_B T)^2}{\pi\hbar^2 v_F^2} \mathrm{Li}_2\left(-\exp\left(\frac{e\varphi + E_F - E_C}{k_B T}\right)\right) \approx -\frac{2(k_B T)^2}{\pi\hbar^2 v_F^2} \mathrm{Li}_2\left(-\exp\left(\frac{e\varphi + E_F}{k_B T}\right)\right). \quad \text{(A.1a)}$$

The 2D concentration of holes in the valence band is

$$p_S(\varphi) = \int_0^\infty d\varepsilon \cdot g_p(\varepsilon) f(\varepsilon - E_V + E_F + e\varphi) \equiv \int_0^\infty \frac{2\varepsilon d\varepsilon/(\pi\hbar^2 v_F^2)}{1 + \exp((\varepsilon - E_V + E_F + e\varphi)/k_B T)}$$
$$= -\frac{2(k_B T)^2}{\pi\hbar^2 v_F^2} \mathrm{Li}_2\left(-\exp\left(\frac{-e\varphi - E_F + E_V}{k_B T}\right)\right) \approx -\frac{2(k_B T)^2}{\pi\hbar^2 v_F^2} \mathrm{Li}_2\left(-\exp\left(-\frac{e\varphi + E_F}{k_B T}\right)\right). \quad \text{(A.1b)}$$

The polylogarithm function $\mathrm{Li}_n(z) = \sum_{m=1}^\infty \frac{z^m}{k^n}$ and $E_C = E_V = 0$ for pure planar graphene sheet.

Introducing a new variable, $\psi = \dfrac{e\varphi + E_F}{k_B T}$, the total charge density is

$$\rho_G(\psi) = e(p_S(\psi) - n_S(\psi)) = \frac{2(k_B T)^2 e}{\pi\hbar^2 v_F^2}(\mathrm{Li}_2(-\exp(\psi)) - \mathrm{Li}_2(-\exp(-\psi)))$$
$$= \frac{2(k_B T)^2 e}{\pi\hbar^2 v_F^2} \sum_{m=1}^\infty \frac{(-1)^m}{m^2}\left((\exp(\psi))^m - (\exp(-\psi))^m\right) \quad \text{(A.3)}$$
$$\approx \frac{2(k_B T)^2 e}{\pi\hbar^2 v_F^2}\begin{cases} \left(-\dfrac{\psi^3}{12} - 2\psi \ln(\psi)\right), & |\psi| \ll 1, \\ -\dfrac{\psi^2}{2}, & |\psi| \gg 1 \end{cases}$$

Corresponding Pade approximation

$$\rho_G(\psi) = \frac{2(k_B T)^2 e}{\pi\hbar^2 v_F^2}\left(\frac{1}{\exp(\psi) + \dfrac{2}{\psi^2 + \dfrac{\psi}{2} + \dfrac{2\pi^2}{12-\pi^2}}} - \frac{1}{\exp(-\psi) + \dfrac{2}{\psi^2 - \dfrac{\psi}{2} + \dfrac{2\pi^2}{12-\pi^2}}}\right) \quad \text{(A.4)}$$

Far from the domain wall plane $n_S(\varphi) \approx \int_0^{E_F+e\varphi} \frac{2EdE}{\pi\hbar^2 v_F^2} \equiv \frac{(E_F+e\varphi)^2}{\pi\hbar^2 v_F^2}$ and 

$p_S(\varphi) \approx \int_0^{-E_F-e\varphi} \frac{2EdE}{\pi\hbar^2 v_F^2} \equiv \frac{(E_F+e\varphi)^2}{\pi\hbar^2 v_F^2}$. So that $\rho_G(\varphi) = 0$ as anticipated for the system electroneutrality far from the domain wall plane.

**APPENDIX B. The "fictional" boundary conditions at the source and drain planes**

Contrary to the popular belief that the controlled motion of a solitary domain wall (or a small number of them) in a ferroelectric film under a channel is relatively easy to carry out, this task is hardly feasible for thin films whose lateral surfaces $x = \pm L_{FE}/2$ (designated "real Bcs" in **Fig.1 (b)**) border with air or a solid. First of all, because these surfaces are themselves topological defects, on which a partial or complete pinning of the domain walls occurs, these walls can be "flipped" only hypothetically by applying a very large electrical voltage to the gate that is rather difficult in reality because of the electric breakdown of a thin oxide dielectric layer, as well as of a thin ferroelectric film. In addition, large fields in thin films do not agree with the requirements for energy saving and miniaturization of heterostructures. In a real thin ferroelectric film, there are many domain stripes that expand or contract with the film polarization reversal, rather than a "domain" running along the film in a transverse direction. Thus the motion, birth or annihilation of domain walls inside the film are likely, at that the polarization values on the surfaces $x = \pm L_{FE}/2$ are determined by the screening conditions and pinning centers. In this case, the domain walls naturally move along the channel, but sometimes towards each other.

In this case, it is reasonable to assume general boundary conditions of the third kind for polarization, $\left. P_3 \pm \lambda_\pm^F \frac{\partial P_3}{\partial x} \right|_{x=\pm\frac{L_{FE}}{2}} = 0$, with different extrapolation lengths $\lambda_\pm^F$ at different lateral surfaces of the film $x = +L_{FE}/2$ and $x = -L_{FE}/2$. In general, $\lambda_\pm^F$ are unknown, but, as a rule, regarded not negative (including special cases of zero and infinitely large value). For any positive $\lambda_\pm^F$, it turns out that the domains expand and contract more often when applying periodic stress to the film, rather than on it, a solitary domain wall in the transverse direction. The equilibrium period of the domain structure is practically independent of the boundary conditions, but is determined by the film thickness $h_f$, temperature, and screening conditions on its polar surface $z = h_{DL}$. Therefore, in the realistic case $L_{FE} \gg L$, the solution for the polarization of the film "under" the channel (i.e., at $-L/2 \le x \le L/2$) practically does not depend on the specific type of BCs at the surfaces $x = \pm L_{FE}/2$, but the average conductivity of the channel naturally depends on the commensurability of the channel length and the period of the domain structure. Taking into account that $L_{FE} \gg L$, it is possible to perform calculations on a computational cell with a transverse dimension $-L/2 \le x \le L/2$.

Since the lateral dimension of a ferroelectric film $L_{FE}$ is typically much higher than the graphene channel length $L$, odd, even or not fractional number of domain walls can pass along the channel during the period of the gate voltage $T_g$ depending on the interrelation between the graphene channel length $L$ and the period $T_{FE}$ of the domain structure in a ferroelectric film [**Fig.1(b)**]. The realistic situations can be modelled by a "fictional" boundary conditions (BCs) on polarization and electric potential at the lateral boundaries of our computation box, $x = \pm L/2$, which will be described below.

The *periodic*, *antiperiodic* or *mixed parity* BCs imposed on polarization and electric potential in transverse x-direction model the appearance and motion of different parity (*even* or *odd*) of pnJs along the channel of length $L$ over a voltage period, respectively. Appeared that periodic BCs for polarization and potential,

$$P_3\left(-\frac{L}{2}, y, z\right) = P_3\left(+\frac{L}{2}, y, z\right), \qquad \left.\frac{\partial P_3}{\partial x}\right|_{x=-\frac{L}{2}} = \left.\frac{\partial P_3}{\partial x}\right|_{x=+\frac{L}{2}}, \quad \text{(periodic)} \quad \text{(B.1a)}$$

$$\varphi_f\left(-\frac{L}{2}, y, z\right) = \varphi_f\left(+\frac{L}{2}, y, z\right), \qquad \left.\frac{\partial \varphi_f}{\partial x}\right|_{x=-\frac{L}{2}} = \left.\frac{\partial \varphi_f}{\partial x}\right|_{x=+\frac{L}{2}}, \quad \text{(periodic)} \quad \text{(B.1b)}$$

allow the motion of the even number of domain walls in the ferroelectric and so the even number of pnJs are moving in the channel. Thus the asymmetry of the graphene channel conductance and rectification effect are absent in the case. In contrast the mixed parity BCs,

$$P_3\left(-\frac{L}{2}, y, z\right) = P_3\left(+\frac{L}{2}, y, z\right), \qquad \left.\frac{\partial P_3}{\partial x}\right|_{x=-\frac{L}{2}} = \left.\frac{\partial P_3}{\partial x}\right|_{x=+\frac{L}{2}} \quad \text{(periodic)} \quad \text{(B.2a)}$$

$$\varphi_f\left(-\frac{L}{2}, y, z\right) = -\varphi_f\left(+\frac{L}{2}, y, z\right), \qquad \left.\frac{\partial \varphi_f}{\partial x}\right|_{x=-\frac{L}{2}} = -\left.\frac{\partial \varphi_f}{\partial x}\right|_{x=+\frac{L}{2}} \quad \text{(antiperiodic)} \quad \text{(B.2b)}$$

can lead to the asymmetry of the graphene channel conductance and rectification effect, which will be demonstrated in the next section. The completely antiperiodic BCs:

$$P_3\left(-\frac{L}{2}, y, z\right) = -P_3\left(+\frac{L}{2}, y, z\right), \qquad \left.\frac{\partial P_3}{\partial x}\right|_{x=-\frac{L}{2}} = -\left.\frac{\partial P_3}{\partial x}\right|_{x=+\frac{L}{2}} \quad \text{(antiperiodic)} \quad \text{(B.3a)}$$

$$\varphi_f\left(-\frac{L}{2}, y, z\right) = -\varphi_f\left(+\frac{L}{2}, y, z\right), \qquad \left.\frac{\partial \varphi_f}{\partial x}\right|_{x=-\frac{L}{2}} = -\left.\frac{\partial \varphi_f}{\partial x}\right|_{x=+\frac{L}{2}} \quad \text{(antiperiodic)} \quad \text{(B.3b)}$$

can lead to the motion of the odd number of domain walls in the ferroelectric and hence the odd number pnJs can move in the channel. The asymmetries of the graphene channel conductance and rectification effect are possible in the case.

**APPENDIX C. Derivation of the electric fields in a multi-layer system in homogeneous case**

The problem geometry shown in the **Fig. 1(a)** the system of electrostatic equations acquires the form:

$$\Delta\varphi_o = 0, \quad \text{for} \quad -h_O < z < 0, \quad \text{(oxide dielectric layer "}O\text{")} \quad \text{(C.1a)}$$

$$\Delta\varphi_d = 0, \quad \text{for} \quad 0 < z < h_{DL}, \quad \text{(dead layer "}d\text{")} \quad \text{(C.1b)}$$

$$\left(\varepsilon_{33}^b \frac{\partial^2}{\partial z^2} + \varepsilon_{11}^f \Delta_\perp\right)\varphi_f = \frac{1}{\varepsilon_0}\frac{\partial P_3^f}{\partial z}, \quad \text{for} \quad h_{DL} < z < h_{DL} + h_F. \quad \text{(ferroelectric "}f\text{")} \quad \text{(C.1c)}$$

3D-Laplace operator is $\Delta$, 2D-Laplace operator is $\Delta_\perp$. Boundary conditions to the system (C.1) are fixed potential at the top ($z = -h_O$) and bottom ($z = h_{DL} + h_F$) gate electrodes; the continuity of the electric potential at the graphene layer ($z = 0$) and the equivalence of difference of the electric displacement normal components, $D_3^O = \varepsilon_0\varepsilon_O E_3$ and $D_3^d = \varepsilon_0\varepsilon_d E_3$, to the surface charges in graphene $\sigma_G(x, y)$; and the continuity of the displacement normal components, $D_3^f = \varepsilon_0\varepsilon_{33}^b E_3 + P_3^f$ and $D_3^d = \varepsilon_0\varepsilon_d E_3$, at gap/ferroelectric interface. Explicit form of the boundary conditions (**BCs**) in z–direction is:

$$\varphi_o(x, y, -h_O) = U(t), \quad \text{(C.2a)}$$

$$\varphi_O(x, y, 0) = \varphi_d(x, y, 0), \quad D_3^o(x, y, 0) - D_3^d(x, y, 0) = \sigma_G(x, y), \quad \text{(C.2b)}$$

$$\varphi_d(x, y, h_{DL}) = \varphi_f(x, y, h_{DL}), \quad D_3^d(x, y, h_{DL}) - D_3^f(x, y, h_{DL}) = 0, \quad \text{(C.2c)}$$

$$\varphi_f(x, y, h_{DL} + h_F) = 0. \quad \text{(C.2d)}$$

Let us consider 1D case, also we suppose that distribution is almost homogeneous $P_3 \approx \langle P_3 \rangle$ (which is valid for $\lambda_\pm \gg \sqrt{\varepsilon_0\varepsilon_{33}^b g}$ with the latter length-scale being very small). One has solution for potential distribution:

$$\varphi_O = U - E_O(z - h_{DL} - h_O), \quad \varphi_{DL} = \Psi - z E_{DL}, \quad \varphi_f = -(h_f + z)E_f \quad \text{(C.3)}$$

Here we introduced the designations $E_O$, $E_{DL}$ and $E_f$ as an electric fields in the upper oxide layer, lower dead layer and ferroelectric film as well. Using the conditions of potentials continuity at the interfaces, one could express the electric field via interface potentials $\Phi$ and $\Psi$ as follows:

$$E_O = \frac{\Phi - U}{h_O}, \quad E_{DL} = \frac{\Psi - \Phi}{h_{DL}}, \quad E_f = \frac{-\Psi}{h_f}. \quad \text{(C.4)}$$

The interface potentials $\Phi$ and $\Psi$ should be determined from the conditions of displacement continuity:

$$\varepsilon_0 \varepsilon_{DL} \frac{\Psi - \Phi}{h_{DL}} = \langle P_3 \rangle + \varepsilon_0 \varepsilon_b \frac{-\Psi}{h_f}, \qquad \varepsilon_0 \varepsilon_O \frac{\Phi - U}{h_O} - \varepsilon_0 \varepsilon_{DL} \frac{\Psi - \Phi}{h_{DL}} = \sigma(\Phi) \qquad (C.5)$$

Averaging of the LGD equation (3a) and application of the boundary conditions (3b) gives

$$h_f \langle a_3 P_3 + a_{33} P_3^3 + a_{333} P_3^5 \rangle + g \left( \frac{P_3}{\Lambda_+} \bigg|_{z=0} + \frac{P_3}{\Lambda_-} \bigg|_{z=h_f} \right) = h_f E_f \text{ and finally}$$

$$\left( a_3 + \frac{g}{h_f} \left( \frac{1}{\Lambda_+} + \frac{1}{\Lambda_-} \right) \right) \langle P_3 \rangle + a_{33} \langle P_3 \rangle^3 + a_{333} \langle P_3 \rangle^5 \approx -\frac{\Psi}{h_f} \qquad (C.6)$$

Finally, the system of nonlinear equations (C.4)-(C.6) for $\langle P_3 \rangle$, $\Phi$ and $\Psi$ could be reduced to the following

$$\Phi = \frac{h_{DL}}{\varepsilon_{DL}} \left( -\frac{\langle P_3 \rangle}{\varepsilon_0} + \left( \frac{\varepsilon_b}{h_f} + \frac{\varepsilon_{DL}}{h_{DL}} \right) \Psi \right), \qquad (C.7a)$$

$$\left( \frac{\varepsilon_b}{h_f} + \frac{\varepsilon_O}{h_O} + \frac{\varepsilon_O}{h_O} \frac{\varepsilon_b}{h_f} \frac{h_{DL}}{\varepsilon_{DL}} \right) \Psi = \frac{\varepsilon_O}{h_O} U + \left( \frac{\varepsilon_O}{h_O} \frac{h_{DL}}{\varepsilon_{DL}} + 1 \right) \frac{\langle P_3 \rangle}{\varepsilon_0} + \frac{\sigma(\Phi)}{\varepsilon_0}, \qquad (C.7b)$$

$$U = \left( 1 + \frac{\varepsilon_b}{h_f} \left( \frac{h_{DL}}{\varepsilon_{DL}} + \frac{h_O}{\varepsilon_O} \right) \right) \Psi - \left( \frac{h_{DL}}{\varepsilon_{DL}} + \frac{h_O}{\varepsilon_O} \right) \frac{\langle P_3 \rangle}{\varepsilon_0} - \frac{h_O}{\varepsilon_O} \frac{\sigma(\Phi)}{\varepsilon_0}. \qquad (C.7c)$$

These equations allow us to control the value of electric fields in the oxide and dead layers, since they should be much smaller than the threshold breakdown fields, which are less than 1 V/nm for oxide dielectric (e.g. SiO2) and less than 0.1 V/nm for a paraelectric dead layer.

**Figure S1** shows the polarization relaxation time $\tau_{LK} = \Gamma/|a_3|$ in comparison with the graphene charge relaxation time $\tau_M = \varepsilon \varepsilon_0 / e \eta n_S$ ($\varepsilon_0$ is a universal dielectric constant, $\varepsilon$ is a relative dielectric permittivity, $\eta$ is a mobility of carriers in graphene). The inequality $\tau_{LK} \ll \tau_M$ always works in the vicinity of ferroelectric phase transition (e.g. at $0.9 T_C < T < 1.1 T_C$), where $\tau_{LK}$ diverges due to the critical slowing down effect.

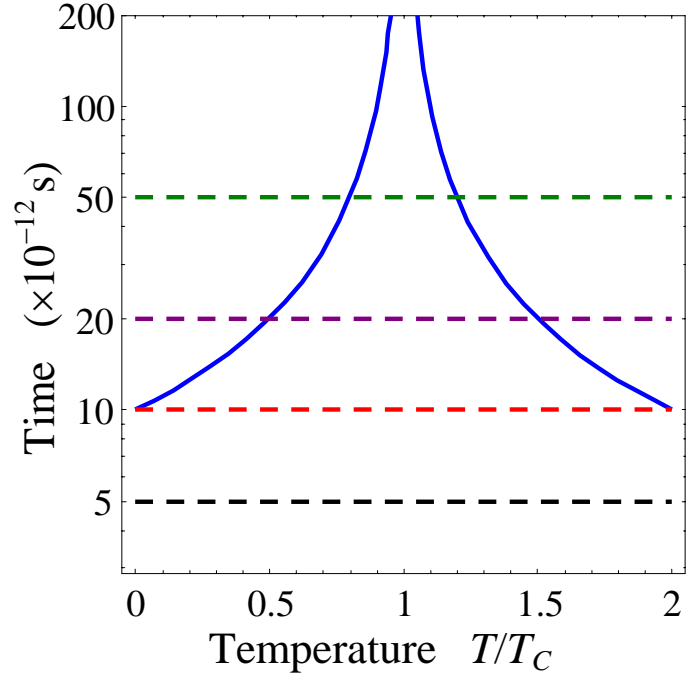

**FIG.S1.** Temperature dependence of the Landau-Khalatnikov relaxation time $\tau_{LK} = \tau_{LK}^0 / |1 - T/T_C|$ (solid curve) in comparison with different charge relaxation time $\tau_M = \varepsilon\varepsilon_0/(e\eta n_S) = 5\times10^{-12}$ s, $10^{-11}$ s, $2\times10^{-11}$ s, $5\times10^{-11}$ s (dashed curves) calculated for the realistic parameter $\tau_{LK}^0 = \Gamma/(\alpha_T T_C) = 10^{-11}$ s.

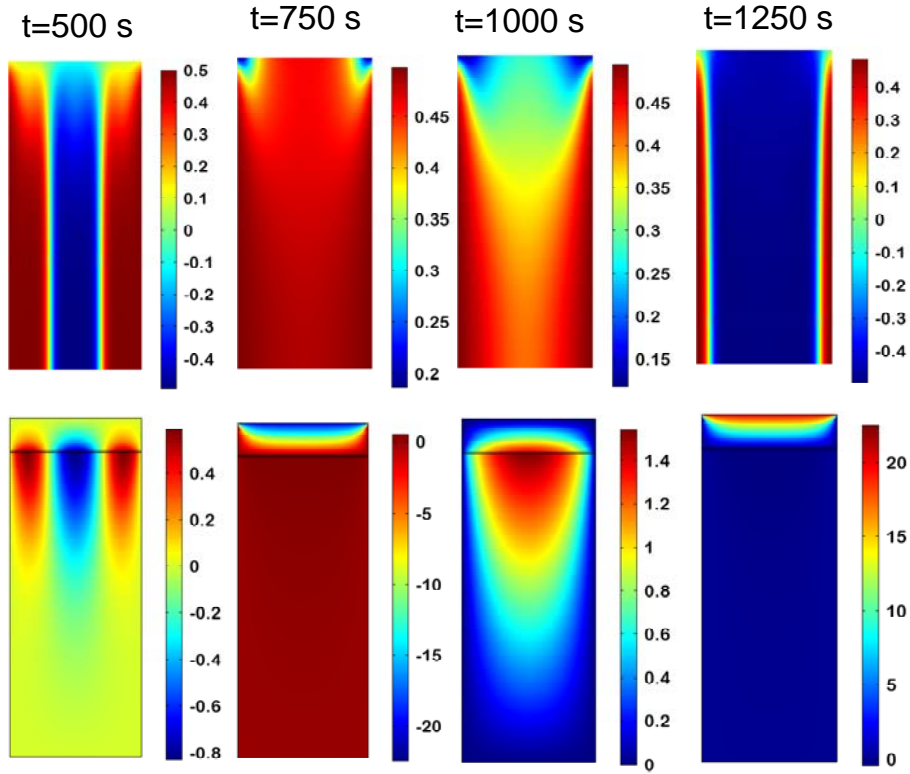

**FIG. S2.** Spatial distribution of domain structure in ferroelectric film (**top row**) and electric potential (**bottom row**) calculated for the mixed BCs (7) at $U_{\max} = 20$ V for different moments of time (500, 750, 1000 and 1250

s). The thicknesses of ferroelectric film is $h_f$=75 nm, oxide dielectric thickness $h_O$=8 nm, dead layer thickness $h_{DL}$=0.4 nm, gate voltage period $T_g$=10³ s. Other parameters are listed in **Table I.**

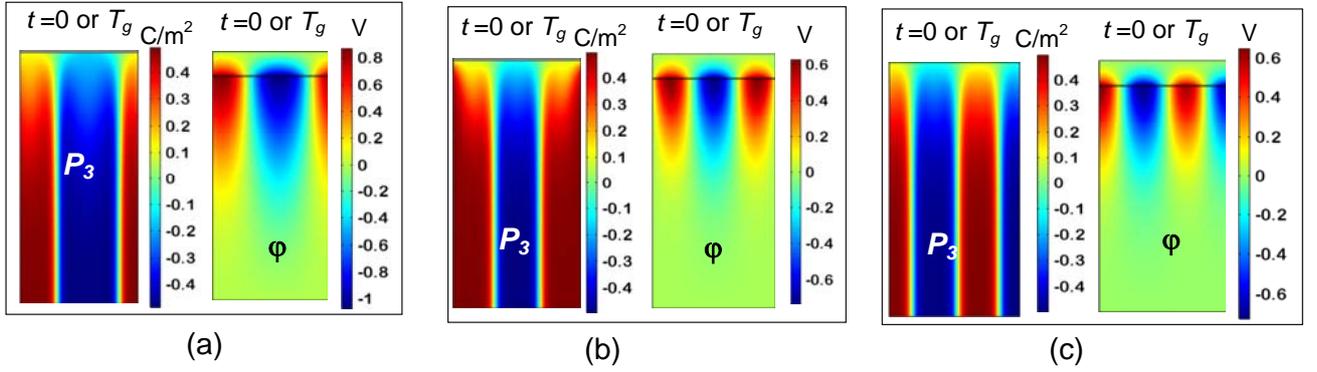

**FIG. S3.** Spatial distribution of polarization component $P_3$ in a ferroelectric film and electric potential φ calculated at t=0 (or Tg), $U_{max}$ = 10 V for the periodic **(a)**, mixed **(b)** and antiperiodic **(c)** boundary conditions (BCs).

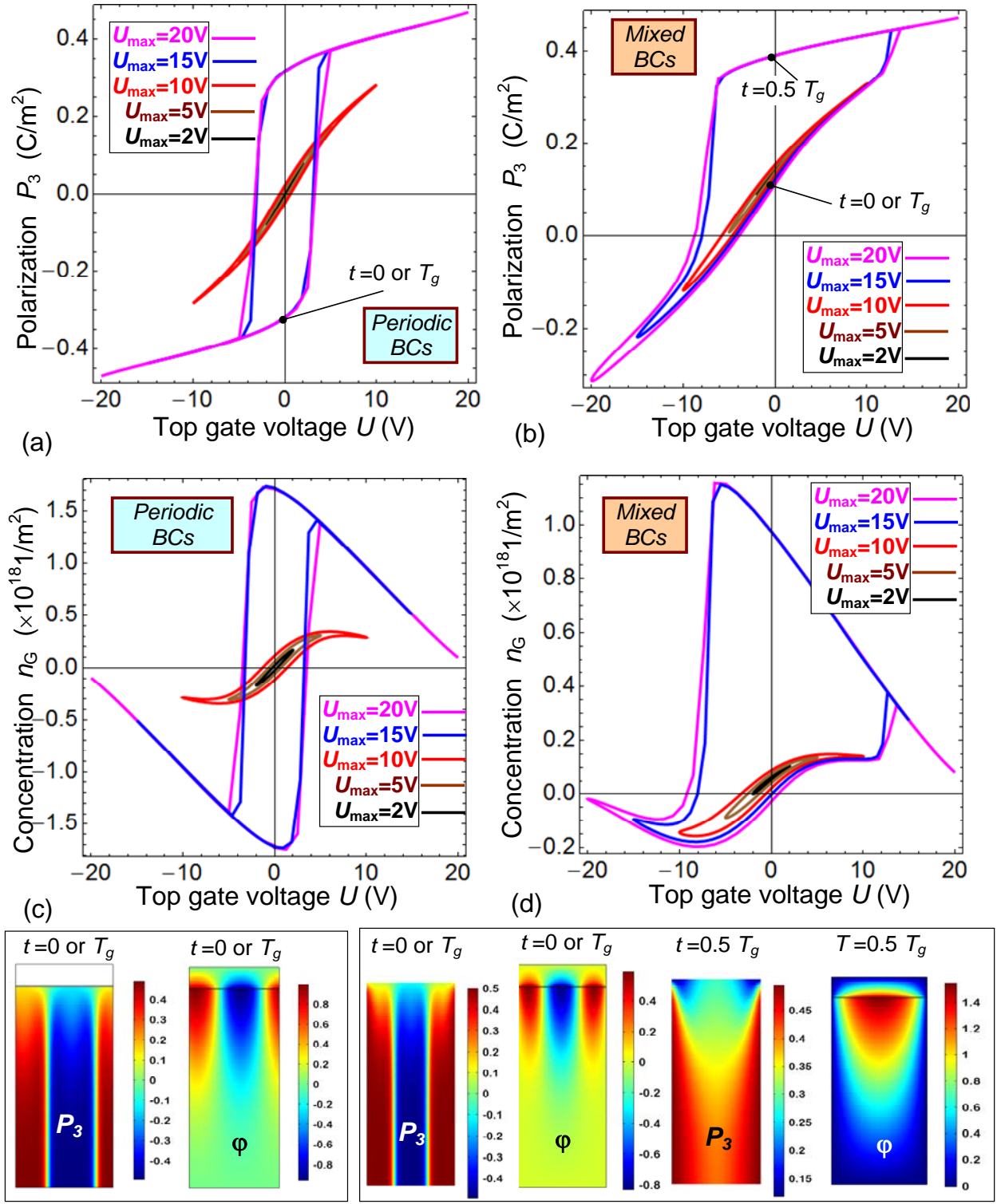

**FIG. S4.** Hysteresis loops of ferroelectric polarization **(a, b)** and surface charge concentration in graphene channel **(c,d)** calculated in dependence on the gate voltage for even **(a, c)** and odd **(b,d)** number of domain walls moving in the ferroelectric substrate. Black, brown, red, blue and magenta loops correspond to the different amplitudes of gate voltage $U_{max}$ = (2, 5, 10, 15, 20) V. Spatial distribution of ferroelectric polarization $P_3$ in a ferroelectric film and electric potential φ calculated for the periodic BCs at $U_{max}$ = 10 V **(left bottom rectangle)** and for mixed BCs at $U_{max}$ = 20 V **(right bottom rectangle)** for different moments of period (0, 0.5 $T_g$ and $T_g$). The thicknesses of ferroelectric film is $h_f$=75 nm, oxide dielectric thickness $h_O$=8 nm, dead layer thickness $h_{DL}$=0.4 nm, gate voltage period $T_g$=10$^3$ s. Other parameters are listed in **Table I**.

9